\documentclass[]{aa}  
\usepackage{subfigure}
\usepackage{xcolor}

\usepackage{graphicx}
\usepackage{txfonts}
\usepackage{natbib,twoopt}
\usepackage[breaklinks=true]{hyperref} 
\bibpunct{(}{)}{;}{a}{}{,}             
\makeatletter
  \newcommandtwoopt{\citeads}[3][][]{\href{http://adsabs.harvard.edu/abs/#3}%
    {\def\hyper@linkstart##1##2{}%
     \let\hyper@linkend\@empty\citealp[#1][#2]{#3}}}
  \newcommandtwoopt{\citepads}[3][][]{\href{http://adsabs.harvard.edu/abs/#3}%
    {\def\hyper@linkstart##1##2{}%
     \let\hyper@linkend\@empty\citep[#1][#2]{#3}}}
  \newcommandtwoopt{\citetads}[3][][]{\href{http://adsabs.harvard.edu/abs/#3}%
    {\def\hyper@linkstart##1##2{}%
     \let\hyper@linkend\@empty\citet[#1][#2]{#3}}}
  \newcommandtwoopt{\citeyearads}[3][][]%
    {\href{http://adsabs.harvard.edu/abs/#3}
    {\def\hyper@linkstart##1##2{}%
     \let\hyper@linkend\@empty\citeyear[#1][#2]{#3}}}
\makeatother

\begin{document}

   \title{Detecting clusters and groups of galaxies populating the local Universe in large optical spectroscopic surveys}

   \author{I. Marini
          \inst{1}
          \and
          P. Popesso\inst{1}$^,$\inst{2}
          \and
          K. Dolag\inst{3}$^,$\inst{4}$^,$\inst{2}
          \and
          M. Bravo\inst{5}
          \and
          A. Robotham\inst{6}$^,$\inst{7}
          \and
          E. Tempel\inst{8}$^,$\inst{9}
          \and
          Q. Li\inst{10}
          \and
          X. Yang\inst{10}$^,$\inst{11}$^,$\inst{12}
          \and  
          B. Csizi\inst{13}
          \and\\
          P. Behroozi\inst{14}$^,$\inst{15}
          \and
          V. Biffi\inst{16}$^,$\inst{17}
          \and 
          A. Biviano\inst{16}$^,$\inst{17}
          \and 
          G. Lamer\inst{18}
          \and
          N. Malavasi\inst{19}
          \and 
          D. Mazengo\inst{1}$^,$\inst{20}
          \and 
          V. Toptun\inst{1}}

   \institute{European Southern Observatory, Karl Schwarzschildstrasse 2, 85748, Garching bei M\"unchen, Germany\\
                \email{ilaria.marini@eso.org}
         \and
            Excellence Cluster ORIGINS, Boltzmannstr. 2, D-85748 Garching bei M\"unchen, Germany
        \and
            Universitäts-Sternwarte, Fakultät für Physik, Ludwig-Maximilians-Universität, Scheinerstr.1, 81679 München, Germany
        \and 
            Max-Planck-Institut für Astrophysik, Karl-Schwarzschildstr. 1, 85741 Garching bei M\"unchen, Germany
        \and
            Department of Physics \& Astronomy, McMaster University, 1280 Main Street W, Hamilton, ON, L8S 4M1, Canada
        \and
            International Centre for Radio Astronomy Research, University of Western Australia, M468, 35 Stirling Highway, Perth, WA 6009, Australia
        \and
            ARC Centre of Excellence for All Sky Astrophysics in 3 Dimensions (ASTRO 3D), Australia
        \and
            Tartu Observatory, University of Tartu, Observatooriumi 1, Tõravere 61602, Estonia
        \and
            Estonian Academy of Sciences, Kohtu 6, 10130 Tallinn, Estonia
        \and 
            Department of Astronomy, School of Physics and Astronomy, Shanghai Jiao Tong University, Shanghai 200240, China
        \and         
            Tsung-Dao Lee Institute, Shanghai Jiao Tong University, Shanghai 200240, China 
        \and    
            Key Laboratory for Particle Astrophysics and Cosmology (MOE) \& Shanghai Key Laboratory for Particle Physics and Cosmology, Shanghai Jiao Tong University, Shanghai 200240, China
        \and
            Universität Innsbruck, Institut für Astro- und Teilchenphysik, Technikerstr. 25/8, 6020 Innsbruck, Austria
        \and
            Department of Astronomy and Steward Observatory, University of Arizona, Tucson, AZ 85721, USA
        \and
            Division of Science, National Astronomical Observatory of Japan, 2-21-1 Osawa, Mitaka, Tokyo 181-8588, Japan
        \and
            INAF – Osservatorio Astronomico di Trieste, Via Tiepolo 11, 34143 Trieste, Italy
        \and 
            IFPU – Institute for Fundamental Physics of the Universe, Via Beirut 2, I-34014 Trieste, Italy
        \and       
            Leibniz-Institut für Astrophysik Potsdam (AIP), An der Sternwarte 16, 14482 Potsdam, Germany
        \and
            Max Planck Institute for Extraterrestrial Physics, Giessenbachstrasse 1, 85748 Garching, Germany
        \and
            Department of Physics, College of Natural and Mathematical Sciences, The University of Dodoma, P.O. Box 338 Dodoma, Tanzania
        }

   \date{Received  ; accepted  }

   \abstract
    {With the advent of wide-field cosmological surveys, samples of hundreds of thousands of spectroscopically confirmed galaxy groups and clusters are becoming available. While these large datasets offer a valuable tool to trace the baryonic matter distribution, controlling systematics in identifying host dark matter halos and estimating their properties remains crucial.}
    {We intend to evaluate the predictions on retrieving the cluster and group of galaxies population using three group detection methods on a simulated dataset replicating the GAMA selection. Our goal is to understand the systematics and selection effects of each group finder, which will be instrumental for interpreting the unprecedented volume of spectroscopic data from SDSS, GAMA, DESI, and WAVES, and for leveraging optical catalogues in the (X-ray) eROSITA era to quantify the baryonic mass in galaxy groups.}
    {We simulate a spectroscopic galaxy survey in the local Universe (down to $z<0.2$ and stellar mass completeness $M_{\star}\geq10^{9.8} M_{\odot}$) using a lightcone based on the cosmological hydrodynamical simulation Magneticum. We assess the completeness and contamination levels of the reconstructed halo catalogues and analyse the reconstructed membership. Finally, we evaluate the halo mass recovery rate of the group finders and explore potential improvements.}
    {All three group finders demonstrate high completeness ($>80$\%) at the galaxy group and cluster scales, confirming that optical selection is suitable for probing dense regions in the Universe. Contamination at the low-mass end ($M_{200}<10^{13} M_{\odot}$) is caused by interlopers and fragmentation. Galaxy membership is at least 70\% accurate above the group mass scale; however, inaccuracies can lead to systematic biases in halo mass determination using the velocity dispersion of galaxy members. We recommend using other halo mass proxies less affected by contamination, such as total stellar luminosity or mass, to recover accurate halo masses. Further analysis of the cumulative luminosity function of the galaxy members has shown remarkable accuracy in the group finders' predictions of the galaxy population.}
    {These results confirm the reliability and completeness of the spectroscopic catalogues compiled by these state-of-the-art group finders. This paves the way for studies that require large sets of spectroscopically confirmed galaxy groups and clusters or studies of galaxy evolution in different environments.}


   \maketitle
%
\section{Introduction}
In the standard cosmological model, galaxies (or visible matter, in general) are bound to deep dark matter (DM) potentials, a dense environment which favours cooling and condensation of baryons, forming cosmological structures \citep{peebles_large-scale_1980, mo_galaxy_2010}. Theoretically, DM halos are the primary drivers of gravitational collapse shaping the evolution of the galaxies they host and their surrounding medium. Numerous observational studies, such as the morphology-density relation \citep[e.g.][]{dressler_galaxy_1980}, the star formation activity or colour-density relations \citep{gomez_galaxy_2003}, and the central galaxy-halo mass relation \citep{moster_constraints_2010, behroozi_average_2013, behroozi_universemachine_2019}, naturally align with the hierarchical paradigm of structure formation.
\par
\begin{figure*}
    \centering
    \includegraphics[scale=0.85, trim={0 5cm 0 4.8cm},clip]{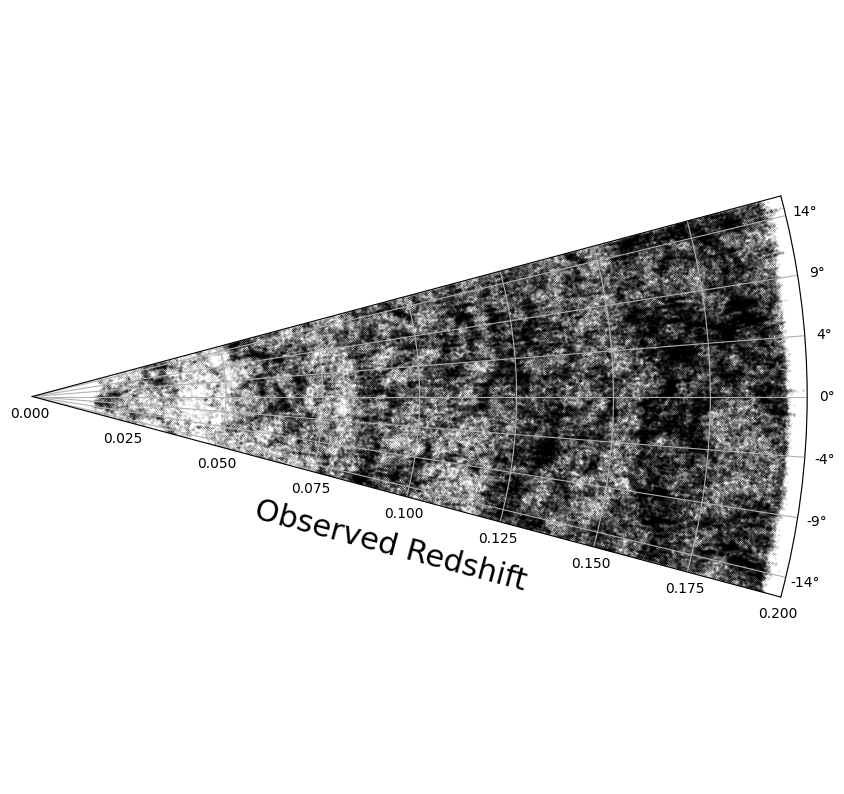}
    \caption{Distribution of the galaxies in LC30. A stellar mass cut $\geq10^{9.8} M_{\odot}$ is applied to the galaxy sample. The survey is depicted in the redshift space down to redshift $z<0.2$.}
    \label{fig:gal_30x30_z_obs}
\end{figure*}
Galaxy groups, the most common galaxy environments \citep{eke_where_2005}, are often difficult to identify. For instance, our own Milky Way, with its close companion and dwarf satellites, forms what we would classify as a typical loose group. Furthermore, halo mass estimates are derived from X-ray or spectroscopic observations. Due to the low X-ray emission and limited member counts in galaxy groups, most studies linking galaxy and gas properties to halo mass focus on extreme environments, such as galaxy clusters. 
\par
The intragroup-medium in low-mass halos emits primarily through Bremsstrahlung radiation or metal line emissions in the X-rays. Still, detecting this gas at temperatures below  $1-2$ keV has been a significant challenge for previous X-ray surveys, which often lacked the sensitivity or coverage needed for these lower mass ranges \citep{ponman_rosat_1996, mulchaey_x-ray_2000, osmond_gems_2004, sun_chandra_2009, lovisari_scaling_2015}. This has led to a substantial "group desert" in main scaling relations, making it difficult to study galaxies in groups unless they are among the brightest in X-rays. However, the launch of eROSITA aboard SRG in 2019 promised to change this. The first catalogue based on the eROSITA All Sky Survey (eRASS) \citep{merloni_srgerosita_2024} and future, deeper data releases covering half the sky are expected to fill this "group desert" in X-ray scaling relations. Nonetheless, deeper analyses from the eROSITA Final Equatorial-Depth Survey (eFEDS) suggest that even at its nominal eRASS:8 depth, eROSITA only detects a small fraction of galaxy groups below halo masses of $10^{14} M_{\odot}$ \citep{popesso_x-ray_2024}. Synthetic eROSITA data, derived from Magneticum hydrodynamical simulation lightcones, indicate that eROSITA's selection function is biased against galaxy groups with lower surface brightness profiles and higher core entropy \citep{marini_detecting_2024}. Consequently, future X-ray large-scale surveys may continue to provide a skewed view of the relationship between galaxies, gas, and their host halo masses.
\par
In parallel with the efforts of the X-ray community, the last few decades have seen an increasing number of large (optical) spectroscopic redshift surveys and continuous development of group finder algorithms to use in these surveys. These algorithms extract the galaxy clustering information from highly complete samples, rather than gas distribution, to identify DM halos and infer their masses. As a result of these large-scale surveys, numerous group catalogues have been constructed, including those from the CfA redshift survey \citep[e.g.][]{geller_large-scale_1987}, the Las Campanas Redshift Survey \citep[e.g.][]{tucker_loose_2000}, the 2dFGRS \citep[e.g.][]{colless_2df_2001, eke_galaxy_2004, yang_halo-based_2005, tago_clusters_2006}, the high-redshift DEEP2 survey \citep[][]{gerke_deep2_2005}, the Two Micron All Sky Redshift Survey \citep[2MASS; e.g.][]{crook_groups_2007, diaz-gimenez_where_2015, lu_galaxy_2016, lim_galaxy_2017}, zCOSMOS \citep{wang_identifying_2020}, and most notably the SDSS \citep{yang_galaxy_2007, tempel_merging_2017}, and GAMA \citep{driver_galaxy_2022, robotham_galaxy_2011}. Various group catalogues based on SDSS observations have been developed using the friends-of-friends (FOF) algorithm \citep[e.g.][]{berlind_percolation_2006, merchan_galaxy_2005}, the C4 algorithm \citep[e.g.][]{miller_c4_2005}, and the halo-based group finder developed by \citet{yang_halo-based_2005} \citep[e.g.][]{weinmann_properties_2006, yang_galaxy_2007, duarte_maggie_2015, rodriguez_combining_2020, yang_extended_2021} and \cite{tempel_friends--friends_2016}. While the SDSS group catalogues are limited to the local Universe, mainly at $z<0.2$, the deeper GAMA survey, with a magnitude limit two magnitudes deeper than the SDSS in the $r$-band and spectroscopic completeness of 95\%, provides a galaxy group sample based on the FOF algorithm of \cite{robotham_galaxy_2011} up to $z\sim0.5$. The deeper magnitude limit and the extremely high spectroscopic completeness enable us to capture the most common galaxy pairs as our own Milky Way and the Andromeda galaxy, thus, sampling the most common galaxy environment.
\par
Nevertheless, concerns persist regarding potential contamination in optically selected group samples and uncertainties in their halo mass measurements. Indeed, while different selection effects might be attributed to different group samples due to the varying depth and selection functions of the spectroscopic galaxy samples on which they are based, large discrepancies might also arise from different approaches to measuring group halo mass \citep[see][]{wojtak_galaxy_2018}. Some algorithms use velocity dispersion-based halo mass estimates \citep[e.g. the GAMA catalogue and the SDSS sample of ][]{tempel_merging_2017}, which heavily depend on the galaxies used to estimate the velocity distribution. Others, such as the algorithms by \cite{yang_cross-correlation_2005} and \cite{tinker_self-calibrating_2021}, rely on different calibrations of the total luminosity or stellar mass-halo mass correlation.
\par
While each of these group finders has been tested on dedicated simulations to compare input and output consistency and estimate uncertainties, a direct comparison of different algorithms based on the same spectroscopic survey is still missing. In the current paper, we aim to benchmark the predictions of three optical group detection algorithms \citep{robotham_galaxy_2011, yang_halo-based_2005, tempel_merging_2017} on the same synthetic dataset which mimics the GAMA selection. These group finders have been (or will be) extensively used on SDSS \citep{yang_galaxy_2007, tempel_flux-_2014, tempel_merging_2017}, GAMA \citep{driver_galaxy_2022}, DESI \citep{desi_collaboration_desi_2016}, DEVILS \citep{davies_deep_2018}, and WAVES \citep{driver_wide_2016} data, providing us with an unprecedented volume of spectroscopic data. Naturally, their efficiency can only be calibrated and tested in controlled experiments, when the properties of the associated DM halos are known, for example in mock observations created with hydrodynamical simulations. The goal is to understand the systematics and selection effects for each of the optical group finders and assess their reliability in their predictions \citep{popesso_perils_2024}. In the eROSITA era \citep{merloni_erosita_2012}, the combination of these optical catalogues will help us shed light on the baryonic mass in groups, which remains a topic of ongoing debate today \citep[see, for example,][]{oppenheimer_simulating_2021}. Recently, optical selection has been used to stack eROSITA data on SDSS \citep{zhang_hot_2024} and GAMA \citep{popesso_x-ray_2024, popesso_average_2024} galaxy groups disclosing the doors to probe the X-ray undetected sources. For this particular experiment, we will focus on the performance of the three algorithms in a GAMA-like survey in the local Universe at $z< 0.2$. However, in the future, we will test the algorithms on a WAVES-like survey at high redshift to check the reliability of the group finders in the more distant Universe. 
\par
The paper is structured as follows. In Sect.~\ref{sec:simulations}, we present the simulation set and the optical lightcone extracted. Sect.~\ref{sec:group_finders} describes the optical halo finders run to detect the galaxy groups and clusters. Sect.~\ref{sec:detecting} illustrates the outcome of the detection procedure evaluating completeness, contamination, and halo mass proxies. Sect.~\ref{sec:hidden_bias} focuses on the optical selection effects and their implications in extragalactic surveys. Finally, Sect.~\ref{sec:conclusions} concludes our study, providing a summary of our findings.

\section{Simulations}
\label{sec:simulations}
\subsection{The Magneticum simulation}
The Magneticum Pathfinder simulation\footnote{\url{http://www.magneticum.org/index.html}} is an extensive series of cosmological hydrodynamical simulations performed using the TreePM/SPH code P-GADGET3. The latter is an improved version of the publicly available GADGET-2 code \citep{springel_cosmological_2005}, while introducing several key advances, such as a higher-order kernel function, time-dependent artificial viscosity, and artificial conduction schemes \citep{dolag_turbulent_2005, beck_improved_2016}.
\par Subgrid models account for the unresolved baryonic physics, including radiative cooling \citep{wiersma_effect_2009}, a time-evolving UV background \citep{haardt_modelling_2001}, star formation, stellar feedback \citep[i.e. galactic winds;][]{springel_cosmological_2003}, and chemical enrichment due to stellar evolution \citep{tornatore_chemical_2007}, explicitly tracking multiple elements (i.e. H, He, C, N, O, Ne, Mg, Si, S, Ca, Fe). They also incorporate models for supermassive black holes (SMBHs) and feedback from active galactic nuclei (AGN), based on the frameworks developed in \cite{springel_cosmological_2005, di_matteo_energy_2005, fabjan_simulating_2010, hirschmann_cosmological_2014}.
\par 
The specific simulation run referenced in this work, known as \textit{Box2/hr}, tracks the evolution of $2 \times 1584^{3}$ particles in a large cosmological volume with dimensions of $(352  h^{-1}$ cMpc$)^{3}$. The particle masses are set at $m_\mathrm{DM} = 6.9\times10^8 h^{-1} M_{\odot}$ for dark matter and $m_\mathrm{gas} = 1.4\times10^8 h^{-1} M_{\odot}$ for gas particles. The softening lengths are $\epsilon=3.75 h^{-1}$ kpc for dark matter, gas, and black hole particles, whereas stars have $\epsilon=2 h^{-1}$ kpc at $z=0$.
\par
Post-processing uses a Friends-of-Friends (FOF) algorithm followed by SubFind \citep{springel_populating_2001, dolag_substructures_2009} to identify halos and substructures (i.e. galaxies). Additional details on SubFind are provided in Sect.~\ref{sec:lightcone}.
\par 
The cosmological parameters used in the simulations follow the WMAP7 values \citep{komatsu_hunting_2010}: $\Omega_\mathrm{M}=0.272$, $\Omega_\mathrm{b}=0.0168$, $n_s=0.963$, $\sigma_8 = 0.809$, and $H_0=100 h$ kms${^{-1}}$~Mpc${^{-1}}$, with $h=0.704$.

\subsection{Designing the lightcone}
\label{sec:lightcone}
The lightcone is extracted from the parent cosmological {\it Box2/hr} described above. Its geometrical design is extensively outlined in \cite{marini_detecting_2024}, where a complementary X-ray catalogue is produced. Here, we provide only the essential information and the differences related to mocking the optical counterpart.
\par
Clusters, groups, and member galaxies are all identified by the halo finder SubFind \citep{springel_populating_2001, dolag_substructures_2009} in the parent box. SubFind is a refined step in the structure identification procedure, after the FOF run. The algorithm descends along the density gradient given by the particles to find the local maxima and minima of the gravitational potential. Each three-dimensional volume encompassed by local minima is a potential substructure (or subhalo) candidate. In addition, SubFind implements an unbinding procedure, to include only particles that are gravitationally bound to each substructure. This step eliminates particles whose internal energy is positive (unbounded particles): if more than a certain minimum number of particles ($50$) survives the unbinding, the substructure is identified as a genuine subhalo. Therefore, every future reference to clusters and groups in the paper will correspond to the identified FOFs (including the central and satellite galaxies); the member galaxies are the sole bound substructures within. The centre of each halo is identified with the minimum gravitational potential occurring among the member particles, other than a complete set of observables (e.g. stellar mass, halo mass, star formation) computed by integrating the properties of the constituent particles. 
\par
 We use the SubFind galaxy and group catalogue as a base to construct the mock galaxy catalogue limited to the local Universe (up to $z < 0.2$) and covers an area of $30\times30$ deg$^2$. The galaxy mock catalogue stores the observer-frame magnitudes in the SDSS filters ({\it u, g, r, i, z}), the observed redshifts, the stellar mass, and the projected position on the sky (i.e. RA, Dec) for each galaxy. Stellar magnitudes \citep[more details in][]{saro_properties_2006} are calculated assuming a single stellar population model for each star particle (with its initial mass, metallicity, and redshift of formation), an initial mass function \citep{chabrier_galactic_2003}, and stellar evolution tracks \citep{girardi_evolutionary_2000}. To accurately account for observational uncertainties and incorporate the K-correction, the rest-frame magnitudes given by Magneticum have been fitted through a standard SED fitting technique with CIGALE \citep[][and references therein]{yang_fitting_2022}. The best-fit template is then shifted to the observed frame at the galaxy's redshift, and the magnitudes in the desired filters are recalculated from this. This approach inherently includes the K-correction. Furthermore, the observed redshift and stellar masses have errors sampled from a Gaussian distribution with $\sigma=45$ km s$^{-1}$ and 0.2 dex $M_{\odot}$, respectively. Additionally, we set 5\% of the galaxies to undergo catastrophic failure in the spectroscopic survey (i.e. $\Delta v > 500$ km s$^{-1}$). This design aims to mock the spectroscopic completeness in GAMA, thus benchmarking our pipelines. No magnitude cut is applied to the sample.
\par
\begin{figure}
    \centering
    \includegraphics[scale=0.75]{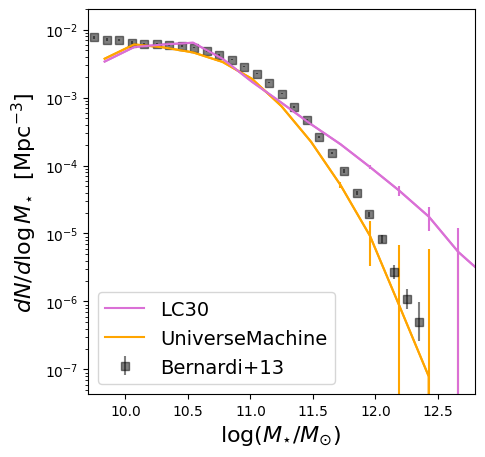}
    \caption{GSMF of the lightcone from Magneticum and UniverseMachine within $z<0.2$. The mock catalogues are compared to the results from the SerExp model in \cite{bernardi_massive_2013}.  }
    \label{fig:GSMF}
\end{figure}

The resulting distribution of galaxies in the redshift space of LC30 is illustrated in Fig.~\ref{fig:gal_30x30_z_obs}. Structures such as filaments and voids are present at all redshifts, indicating a variety of environments in which galaxy groups and clusters reside. Magneticum has provided reliable predictions for its simulated galaxy population \citep[e.g.][]{teklu_connecting_2015, teklu_dynamical_2016, teklu_morphology-density_2017, teklu_decline_2023, remus_outer_2017, schulze_kinematics_2017, schulze_kinematics_2018, schulze_kinematics_2020, popesso_hot_2024}. We limit the galaxy sample to a stellar mass of $\geq 10^{9.8} M_{\odot}$ to ensure completeness for Magneticum's stellar mass resolution. However, the chosen cosmological box \textit{Box2/hr} shows a significant tension at the massive end of the galaxy stellar mass function (GSMF) compared to observational data. In Fig.~\ref{fig:GSMF}, we illustrate this comparison between the GSMF from LC30 (with Poisson uncertainty) and \cite{bernardi_massive_2013}. Galaxies with $\log (M_{\star}/M_{\odot}) > 11.5$ are primarily the central galaxies of the simulated galaxy clusters and groups, leading our simulations to predict more massive BCGs and BGGs compared to observed ones \citep[see also discussion in][]{ragone-figueroa_bcg_2018}. 
\par
The tension between observed and simulated stellar masses at these scales has been addressed in many works \citep[e.g.][]{pillepich_first_2018, bassini_dianoga_2020}. Part of the tension is necessarily linked to how different simulations and observations estimate the stellar mass. Oftentimes, the observed stellar luminosity in a given central aperture is converted with a mass-to-light ratio and extrapolated to larger radii \citep[e.g.][]{kluge_structure_2020}. A less diffuse and more expensive method requires near-infrared luminosity since the K-band is a good measure of its underlying stellar mass regardless of how that mass assembled itself \citep{kauffmann_k-band_1998}. Simulations tend to replicate these results (for instance, by measuring the mass of the star particles within a given central aperture) alleviating such tensions \citep[see][for example]{kravtsov_stellar_2018}, however, an important factor is played by the AGN feedback implementation \citep{ragone-figueroa_brightest_2013} and the uncertainty on the intracluster light contribution \citep[e.g.][]{gonzalez_census_2007, montes_faint_2022}.
\par
This effect might bias our analysis if group finders are systematically led to identify the central galaxy due to the massive end of the GSMF distribution. We assess this effect in Appendix~\ref{sec:UniverseMachine} by performing the same analysis on a lightcone with the same volume but produced with UniverseMachine \citep{behroozi_universemachine_2019, behroozi_universemachine_2023}. We create a galaxy mock catalogue on which we run a group finder to test the hypothesis on the assumed GSMF in Magneticum. This is possible since the UniverseMachine mock catalogue derives from empirical models of galaxy formation, specifically calibrated to the most recent observational data, thus tuned to reproduce observables such as the GSMF. Our investigation shows that the findings in Magneticum are robust and they are not biased by the central galaxy modelling. 
\par

\section{Group finders}
\label{sec:group_finders}
Here, we briefly present the three group finders we used for this experiment. Groups are identified with a geometrical criterion derived by a FOF in all three methods, however, parameter tuning and halo mass estimation vary from one another. Although reconstructing halo masses is a crucial step in the parameter exploration, connecting a given true observable with halo mass is not trivial and many efforts have been put into improving and understanding systematics there \citep[e.g.][]{saro_toward_2013, old_galaxy_2014, old_galaxy_2018, wojtak_galaxy_2018, vazquez-mata_galaxy_2020}. Here, we attempt to address all crucial points in such estimates, progressing into a deeper discussion of different halo mass proxies in Sect.~\ref{sec:mass_proxy}.
\subsection{Tempel et al. (2017)}
The group finder described in \cite{tempel_merging_2017} (T17, hereafter) has been used to determine the group catalogue in the SDSS data release 12. Here, we list the main steps of the algorithm. 
\begin{enumerate}
    \item The galaxy catalogue is run through a FOF algorithm in redshift space. For its nature, the linking lengths are required to be different in the transversal and radial (i.e. along the line of sight) direction. The transversal linking length depends only on the redshift and is calibrated using the mean separation of galaxies in the plane of the sky as described in \cite{tempel_flux-_2014}. Radial linking length is taken to be ten times the transversal linking length.
    \item Next, the algorithm runs a group membership refinement to filter nearby field galaxies or filaments from incorrect group assignments, according to \cite{tempel_friends--friends_2016}. The first step involves multimodality analysis to separate multiple components within groups into distinct systems. This is achieved through a model-based clustering analysis using the \textit{mclust} package in the statistical computing environment R. In this analysis, one coordinate axis is fixed with the line of sight, while the other two are allowed free orientation in the sky plane. This clustering analysis is applied to groups with at least seven galaxies. For each potential number of subgroups (ranging from one to ten), \textit{mclust} determines the most probable locations, sizes, and shapes of these subgroups. The Bayesian information criterion is used to select the number of subgroups, and each galaxy is assigned to a group based on the highest probability calculated by \textit{mclust}.
    \item The second step in membership refinement involves estimating the group’s radius $R_{200}$, assuming a NFW profile. A galaxy is excluded from a group if its distance in the sky plane from the group centre exceeds the virial radius, or if its velocity relative to the group centre exceeds the escape velocity at its projected distance from the group centre. This exclusion process is carried out iteratively and usually converges after a few iterations. The refinement process is applied only to groups with at least five members.
    \item Finally, the group detection and membership refinement procedures are reiterated for all excluded members to determine if they form separate groups. This step is crucial for detecting small groups that may have been missed during the initial multimodality analysis.
\end{enumerate}
The derivation of the group masses is estimated only for groups with three or more members. The group finder uses the virial relation $M_{200}\propto \sigma^2 R_{200}$ which connects the mass $M_{200}$\footnote{We define $M_{\Delta}$ as the mass encompassed by a mean overdensity equal to $\Delta$ times the critical density of the universe $\rho_c(z)$.} to the member's velocity dispersion $\sigma$ and extension $R_{200}$. The group extent in the sky and velocity dispersion are not clearly defined for galaxy pairs. However, the estimated group mass is also largely uncertain for other poor groups. By iteratively estimating the velocity dispersion and extension, one can determine the group's mass. 

\subsection{Yang et al. (2005)}
The group finder described in \cite{yang_halo-based_2005} (Y05, hereafter) has been successfully applied to numerous galaxy samples with both spectroscopic and photometric redshifts \citep[e.g.][]{yang_cross-correlation_2005, yang_galaxy_2007, yang_extended_2021}. Its strengths lie in its iterative nature and adaptive filter modelled after the general properties of DM halos.
Here, we outline the procedure, while the interested reader can find a more detailed description in \cite{yang_halo-based_2005, yang_galaxy_2007, yang_extended_2021}.
\begin{enumerate}
    \item The group finder starts by assuming that all galaxies are tentative groups. Then, abundance matching between the total group luminosity and the halo mass is used to obtain the mass-to-light ratios iteratively. 
    \item To reduce the influence of the survey magnitude limit on the halo mass estimation, different redshift bins are used to divide galaxy samples, and their mass-to-light ratios are determined individually. By interpolating in both redshift and group luminosity, the halo mass for each group is derived. Based on the halo mass, the size and velocity dispersion of the underlying DM halo are computed.
    \item Based on the DM halo properties, the group finder assigns galaxies assuming that the phase-space distribution of galaxy members follows the DM particles, where the luminosity-weighted centre is used to trace the group centre. Therefore, the number density contrast of galaxies in redshift space around the group centre at redshift $z_\mathrm{group}$ can be written as 
    \begin{equation}
        P_M(R, \Delta z) = \frac{H_0}{c}\frac{\Sigma(R)}{\Bar{\rho}} p(\Delta z).
    \end{equation}
    Here, $\Delta z= z-z_\mathrm{group}$, $c$ is the speed of light, $\Bar{\rho}$ is the average density of the universe, and $\Sigma(R)$ is the projected surface density of a spherical NFW halo. The function $p(\Delta z) d\Delta z$ describes the redshift distribution of galaxies, assumed to be Gaussian. 
    \item Galaxies with $P_M \ge B$, where $B=10$ is the background level, are assumed to be the member galaxies of this group. If a galaxy fulfils this criterion for more than one group, it is only assigned to the group with the highest $P_M$. If all members of two groups can be assigned to one group according to the above criterion, the two groups are merged into a single group.
    \item Once all the galaxies are assigned to groups, the group centres are recomputed and iterated to step 2 until group memberships are stable.
\end{enumerate}
Within the above framework, \citet{yang_extended_2021} have extended this algorithm so that it can simultaneously deal with galaxies with either spectroscopic or photometric redshifts. 
The halo masses estimated with this procedure are calculated within the virial definition described in \cite{bryan_statistical_1998}, which uses an overdensity equal to $\Delta = 180$. Therefore, we estimate a correction factor to translate $M_{180}$ into $M_{200}$, assuming a model for the concentration \citep{diemer_colossus_2018}.

\subsection{Robotham et al. (2011)}
The group finder outlined in \cite{robotham_galaxy_2011} (R11, hereafter) has processed the galaxies in the GAMA survey to extract a galaxy group catalogue. The algorithm also relies on a (single-step) FOF procedure to identify groups, with the base choice of radial and projected maximum linking lengths (2 free parameters) being scaled on a per-galaxy basis as a function of both local density contrast (five free parameters) and galaxy brightness (one free parameter).
\par
The free parameters were calibrated against a set of simulated lightcones that reproduced the observed GAMA luminosity function, with only groups with five or more members used to determine the appropriate combination of parameters: five or more members are required to make a meaningful estimate of the dynamical velocity dispersion and 50$^{th}$ percentile radius Rad50.
\par
In this work, we use the same values adopted for the latest version of the GAMA galaxy group catalogue, except for the parameter controlling the link scaling with galaxy brightness which we set to 0 (i.e. no scaling)\footnote{The value adopted in GAMA increases the linking lengths for bright galaxies. With the over-prediction on the number of massive (bright) galaxies in LC30 (shown in Figure \ref{fig:GSMF}), the group finder merged multiple haloes, leading to a significant over-estimation of the high-mass end of the HMF.}.
\par
The halo properties -- such as mass, total luminosity, centre, and velocity dispersion -- are estimated from the galaxy members. The centre is iteratively computed within the brightest galaxies in the group (i.e. IterCen in the catalogue). The velocity dispersion is derived with the gapper estimator \citep{beers_measures_1990} which underweights the outliers. The halo radius considered is Rad50 containing 50 per cent of the galaxies in the group. These two quantities are used to estimate the halo mass from a dynamical argument, since in first order for a virialised system we expect its dynamical mass to scale as $M \propto \sigma^2 R$. \cite{robotham_galaxy_2011} determined the proportionality constant using a semi-empirical estimation from the mock catalogue used to calibrate the fitting parameters, as a function of richness and redshift. The total luminosity needs to account for the missing faint galaxies in the groups, thus, the group finder measures the $r$-band luminosity contained within the survey's limit and then integrates the global galaxy luminosity function to a nominal faint limit used to correct for the missing flux. 
\par
Interestingly, the authors find the group dynamical masses more intrinsically stable (require smaller corrections) as a function of redshift, whilst group luminosities are more stable as a function of multiplicity. Furthermore, the scatter in extrapolated group luminosity is much smaller than seen for dynamical masses \citep{robotham_galaxy_2011}. In this same regard, \cite{vazquez-mata_galaxy_2020} find that using the luminosity estimator as a halo mass proxy returns a better correlation with the true halo mass than the dynamical estimator. We further investigate this issue in Sect.~\ref{sec:mass_proxy}. 
\begin{figure*}
    \centering
    \subfigure{\includegraphics[width=0.44\textwidth]{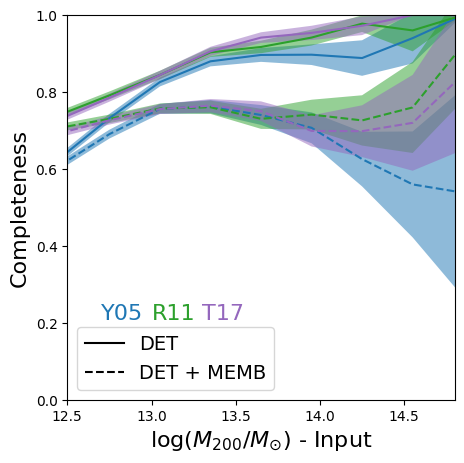}}
    \subfigure{\includegraphics[width=0.46\textwidth]{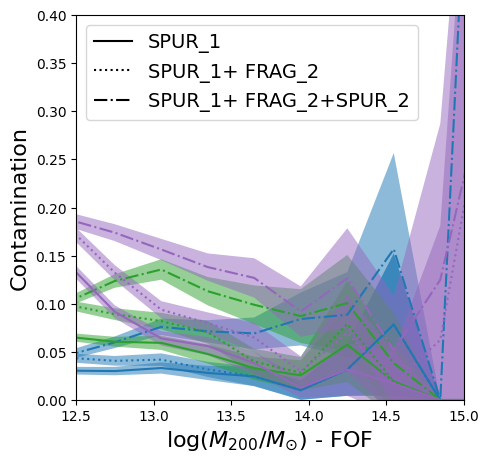}} 
    \caption{Completeness (left panel) and contamination (right panel) as a function of halo mass. The shaded bands mark the binomial confidence interval. The solid lines report the definitions in Eq.~\ref{eq:completness}--\ref{eq:contamination} whereas the dashed lines report stricter ones provided in the main text. Notice that the y-axis is different in the two panels. }
    \label{fig:completeness_contamination}
\end{figure*}

\begin{figure*}
    \centering
    \subfigure{\includegraphics[width=0.45\textwidth]{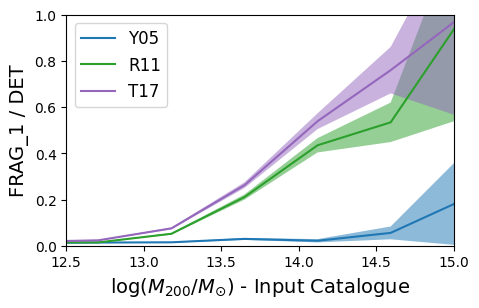}} 
    \subfigure{\includegraphics[width=0.45\textwidth]{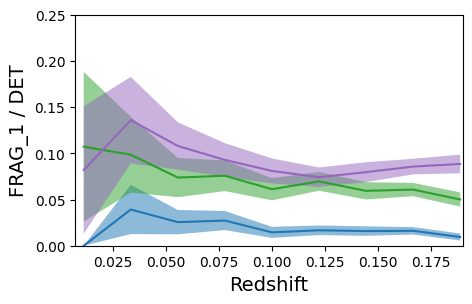}} 
    \caption{Statistical distribution of the fragmentation as a function of the halo mass (left) and redshift (right panel). The shaded bands mark the binomial confidence interval. This represents the fraction of systems that have one or more fragments in the FOF catalogue. The colours follow the same legend used in Fig.~\ref{fig:completeness_contamination}.}
    \label{fig:fragmentation}
\end{figure*}

\section{Detecting optical groups and clusters}
\label{sec:detecting}
\begin{table*}
    \centering
    \begin{tabular}{l|c|c|c|c}
    \vspace{0.1cm}  
 & Magneticum & T17 & Y05 & R11 \\  \hline \hline
    N groups: MEMB>0 & 251,349 & 46,722 & 231,516 &  44,568\\ \hline
    N groups: MEMB>1 & 31,687 & 46,722 & 29,681 &  44,568\\ \hline
    N groups: MEMB>2 & 13,074 & 19,217 & 12,704 &  18,863\\ \hline
    N groups: MEMB>4 & 5,195 & 6,788 & 5,246 &  7,026\\ \hline
    N matched groups:  MEMB>1 & 31,687 & 39,633 & 28,727 &  41,537\\ \hline
    N matched groups: MEMB>1 \& $\log (M_{200}/M_{\odot})>12.5 $ & 17,157 & 17,128 & 13,644 & 15,394\\ \hline
    N matched groups: MEMB>1 \& $\log (M_{200}/M_{\odot})>12.5 $ \\ \hspace{3cm} \& no FRAG\_2 & 17,157 & 14,185 & 14,241 & 15,614\\ \hline
    N matched groups: MEMB>1 \& $\log (M_{200}/M_{\odot})>12.5 $ \\ \hspace{3cm} \& no FRAG\_2 \& no SPUR\_2 & 17,157 & 14,194 & 13,404 & 14,247\\ \hline\end{tabular}
    \caption{Description of the groups in Magneticum and the FOF catalogues. We report all groups in the input halo catalogue (Magneticum) and the detected ones in the FOF catalogues. The following rows record the numbers for the matched groups with a Magneticum group of at least two galaxies, same with a minimum halo mass, excluding FRAG\_2 sources and SPUR\_2.}
    \label{tab:groups}
\end{table*}
Each group finder is tested on LC30 to assess its reliability in recovering the underlying halo population in a blind study. We provide the spectroscopic catalogue containing the five bands' magnitudes, the stellar mass, the observed redshift, and the position in the sky as derived in Sect.~\ref{sec:lightcone}. 
\par In Table~\ref{tab:groups}, we report the 
group numbers from Magneticum and the FOF catalogues, before and after the matching (see next section). The FOF catalogues are based on different definitions of FOF groups: Y05 includes isolated galaxies, whereas R11 and T17 yield groups with a minimum of two members. To ensure equality, we apply a richness cut $\geq2$ galaxies in the FOF catalogue before the matching. 
\subsection{Matching with the input catalogue}
\label{sec:matching}
We evaluate the completeness and contamination of the catalogue by matching it with the input halo catalogue in the mock. In the following, we will refer to quantities with the subscript
\begin{itemize}
    \item "DET" referencing the subsample of true detections;
    \item "HALOS" for all the Magneticum groups and clusters;
    \item "SPUR\_1" for the spurious sources (i.e. unmatched detections);
    \item "FOF" for all the group finders' detections.
\end{itemize}  The matching procedure goes as follows.
\begin{enumerate}
    \item For all the candidate groups in the FOF catalogue (i.e. groups with a minimum of two galaxies, thus no isolated galaxies), we search for a halo counterpart in LC30. We do not apply any richness cut to the Magneticum catalogue, thus isolated galaxies are included in the matching, since a FOF group could be mistakenly associated with them. Notice that this matching would lead to significantly incorrect halo masses; therefore, we treat them as incorrect associations in the contaminants. If a detection is within a maximum offset of $R_{200}$ from the centre, we match it. It is unlikely that the peak stellar emission will be farther away than the virial radius. Additionally, we assess that the estimated redshift is within $3\times 10^{-3}$ of the input redshift of the candidate group. This is comparable to the velocity dispersion of a massive cluster. Thus, it is unlikely that groups identified at larger distances in redshift are associated with the given halo.
    \item If more than one halo falls within the cylindrical volume, we check whether there is one whose mass dominates (i.e. at least six times the second most massive group). If this is the case, we match it; otherwise, we list the primary in the matched catalogue (i.e. the most massive detection), adding the flag FRAG\_1. Many of the later halos "undetected" will be among these secondary detections (with the flag FRAG\_2) and are lost.  
    \item In the end, we check whether, according to these criteria, any halo in the input catalogue has been matched more than once to a FOF detection, and, if so, we define the matching as fragmented.
\end{enumerate}
Using this matching procedure, we classify the detected groups into three categories: primary, secondary, and tertiary detections.
\begin{itemize}
    \item Primary detections: These are either the unique matches or the most massive halos within the detection's $R_{200}$ radius. When fragmentation occurs, the detection with the highest mass is labelled as the primary fragment (FRAG\_1).
    \item Secondary detections: These are fragments of a larger halo split into smaller detections and are labelled as FRAG\_2.
    \item Tertiary detections: These are clearly incorrect matches. For example, we find many two (or more) members FOF groups associated with isolated Magneticum galaxies. This incorrect matching would lead to very biased halo masses. We flag these matches as SPUR\_2.
\end{itemize}
Halos that remain undetected fall into two categories:
\begin{itemize}
    \item Halos obscured by a more massive detection within their $R_{200}$.
    \item Halos without any corresponding FOF detection.
\end{itemize}
These undetected halos are compiled into a single undetected halo catalogue. This categorisation allows us to evaluate the completeness and contamination of our group catalogue by comparing it against the input halo catalogue. 

\subsection{Completeness and contamination}
\label{sec:completeness_contamination}
Here, we discuss the performance of the group finders in detecting groups and clusters of galaxies in our mock observation. We cross-match the detections with Magneticum's halo catalogue; nevertheless, we discuss only the detections corresponding to halos with mass $M_{200}\geq 10^{12.5}\, M_{\odot}$ (i.e. due to the halo completeness limit of our simulation) and with a minimum of two members in the Magneticum catalogue. Therefore, we define completeness as the ratio between the number of detected halos over the total (per input halo mass bin), namely:
\begin{equation}
\label{eq:completness}
    \mathcal{C_\mathrm{ompleteness}} = \frac{N_\mathrm{DET}}{N_\mathrm{HALOS}}.
\end{equation}
It is important to note that this definition does not consider the consistency of galaxy membership in the matching process. Therefore, we redefine completeness as the ratio of detected halos (whose corresponding match includes at least 50 per cent of the member galaxies) to the total. This will be referred to as DET+MEMB.
\par
Similarly, we define contamination as the fraction of spurious detections in the FOF catalogue per bin of FOF halo mass proxy:
\begin{equation}
\label{eq:contamination}
    \mathcal{C_\mathrm{ontamination}} = \frac{N_\mathrm{SPUR}}{N_\mathrm{FOF}}.
\end{equation}
Spurious sources can be SPUR\_1 (i.e. unmatched FOF groups), SPUR\_2 (i.e. incorrect matching), and FRAG\_2 (i.e. fragments of a larger halos). Often in the literature \citep[e.g. in ][]{yang_cross-correlation_2005} we speak of purity (as opposed to contamination) which is simply defined as $\mathcal{P} = 1 - \mathcal{C_\mathrm{ontamination}}$. 
\par
Completeness and contamination are complementary in describing the detection process: completeness defines the accuracy of recovering the input catalogue, whereas contamination clarifies to what extent the FOF catalogue is susceptible to false detections. The optimal case would be completeness equal to 1 and contamination 0, however, there might be cases when both completeness and contamination are high (for example, a group finder detects almost all halos at the cost of detecting extra false halos) or low (equivalently, a group finder cannot find any halo but it does not overpredict halos when there are none). A good group finder should overlook false detections due to improper member classification caused by projection effects or splitting of large groups and clusters. On the other hand, contamination is derived as a function of the recovered halo mass using the estimates from our best halo mass proxy (see Sect.~\ref{sec:mass_proxy}).
\par
Fig.~\ref{fig:completeness_contamination} report our results. In the left panel, we illustrate the completeness of all the primary detections as a function of the true halo mass. For masses above $10^{13}M_{\odot}$ the completeness (solid lines) is at least 80 per cent in all group finders. Alternatively, if completeness is determined solely by the identified halos that have at least 50 per cent of their member galaxies in common with their match (i.e. most of the members overlap), we observe that for masses exceeding $10^{13}M_{\odot}$, the completeness (represented by the dashed lines) is approximately 70 per cent. Galaxy membership is an important step in the group finder, it is also responsible for correctly estimating halo properties, such as halo mass, as discussed in Sect.~\ref{sec:mass_proxy} for example.
\par
In the right panel, we plot the contamination (solid lines) defined in Eq.~\ref{eq:contamination} as a function of the halo mass of the FOF catalogue. This ranges below $10$ per cent for all group finders in the regime $>10^{13}M_{\odot}$. These false detections correspond to groups of a few galaxies, aligned in projection, but not necessarily belonging to a DM halo. Furthermore, a fraction of true detections is fragmented and fragmentation can lead to systematic biases in halo properties. If we include FRAG\_2 as a contributing spurious source in the definition of contamination (dashed lines) the increase is negligible. In the high mass end, contamination is mostly due to SPUR\_2 sources, namely FOF groups that have been associated with Magneticum halos whose mass is 1 dex larger than the true match. These are incorrect matching and a true match is beyond the virial radius used to match.
\par
In Fig.~\ref{fig:fragmentation}, we investigate which primary detections are mostly fragmented. In the left panel, we illustrate the distribution in halo mass for the three group finders. Not surprisingly, most of the fragmented sources are the largest ones. From a more quantitative point of view, the average number of fragments as a function of halo mass is a monotonically increasing distribution for all group finders, reaching four to six fragments at the highest masses. Fragmentation will be almost redshift-independent for the group finders.Since fragmentation occurs in all samples and no direct method can shed light on its nature in a FOF catalogue, we advise care in considering low-mass FOF groups within the projected $R_{200}$ of a large group. 
\par
In conclusion, we argue that optical group finders recover the underlying halo population above $10^{13} M_{\odot}$ with high completeness and low contamination. 

\subsection{Halo mass proxies}
\label{sec:mass_proxy}
A non-trivial task all group finders are required to perform is to recover a reliable halo mass associated with a group of galaxies.
\par
R11 and T17 exploit the virial theorem to connect the velocity dispersion of the member galaxies with $M_{200}$. Using the velocity dispersion as a proxy might lead to inaccurate results for poor groups since a robust velocity dispersion can only be attained with a minimum number of members. In this regard, Y05 argue that a more reliable estimate can be provided by modelling the mass-to-light ratio from the total luminosity of the member galaxies: such a method seems to be only mildly dependent on the assumed mass-to-light ratio. 
\par
\begin{figure}
    \centering
    \includegraphics[scale=0.75]{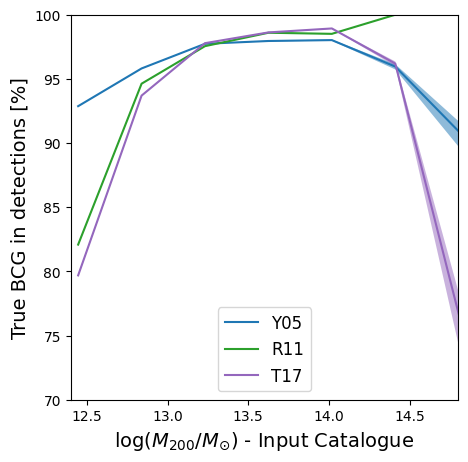}
    \caption{Percentage of BCG correctly identified for each detected halo in the input catalogue. The shaded band marks the dispersion given by the binomial confidence interval. }
    \label{fig:BCG_correct}
\end{figure}

Therefore, we compare the scatter of the distribution when using the halo masses included in the FOF catalogue and, when possible, attempt to improve such estimates with other assumptions. Here, we list the different methods used to reconstruct the halo mass.
\begin{table}
        \centering
        \begin{tabular}{c|c|c}
              \vspace{0.1cm}  $Y$ & $\log M_{200}$ & $\sigma_{\log M_{200, Y}} $ \\ \hline \hline
             $\log M_{\star, BCG}$& 12.5 &  0.197\\
             & 13.0 & 0.200\\
             & 13.5 & 0.200\\
             & 14.0 & 0.180  \vspace{0.1cm} \\ \hline
             $\log M_{\star, TOT}$& 12.5 & 0.170\\
             & 13.0 & 0.145\\
             & 13.5 & 0.126\\
             & 14.0 & 0.117  \vspace{0.1cm} \\ \hline
             $\log L_{r}$ & 12.5 & 0.182\\
             & 13.0 & 0.193\\
             & 13.5 & 0.212\\
             & 14.0 & 0.189 
        \end{tabular}
        \caption{Logarithmic scatter of the scaling relations in Magneticum used as halo mass proxies in the group regime. The proxies $Y$ are reported in the left column and they are the BCG stellar mass, the total stellar mass, and the total luminosity in the $r$-band, respectively. All the masses are in units of $M_{\odot}$, while the luminosity is expressed in $L_{\odot}$.}
        \label{tab:scatter}
\end{table}
\begin{itemize}
    \item \textsc{Stellar mass of the BCG.} Locating the BCG of a galaxy cluster or group is a relatively easy task since it is the brightest (often most massive) galaxy in a virialised system. Its properties are tightly connected to the host DM halo properties, given their co-evolution is predicted to be present since at least $z\leq 2$ \citep{ragone-figueroa_bcg_2018}. Therefore, correctly identifying the central galaxy can help us infer several host halo properties with reasonable uncertainty. The intrinsic scatter in the $M_{\star, BCG}-M_{200}$ relation has been extensively studied \citep[e.g.][]{behroozi_comprehensive_2010, moster_galactic_2013, kravtsov_stellar_2018}. From this, one can conclude that if the algorithms are accurate enough in recovering at least the central galaxy in a group, we can estimate a roughly consistent host halo mass. Fig.~\ref{fig:BCG_correct} reassures us that the consistency of each group finder in recovering the BCG is high, once the matching is done. We report with the shaded band the binomial confidence interval for each sample. Above $10^{13} M_{\odot}$ all algorithms recover at least 90 per cent of all central galaxies, although T17 has a sharp decline at the massive end. 
    In the simulation, the BCG is the most massive galaxy in a halo, generally sitting at the bottom of the gravitational potential. However, the group finders identified the BCG based on its luminosity, being the brightest galaxy in the group. These two definitions in most cases coincide, albeit incorrect galaxy membership and/or equally bright galaxies in a halo might affect the recovery process. We observe that the recovery rate is high in the group regime for all finders overall, decreasing at the low-mass end ($<10^{13} M_{\odot}$ halos formed by one or two galaxies) or the very high-mass end ($>10^{14} M_{\odot}$ where most fragmentation happens). 
    
    \item\textsc{Stellar mass of member galaxies.} Another optical halo mass proxy often used in the literature is the total stellar mass, often estimated by applying a light-to-mass conversion to the galaxy candidate members of a cluster or group \citep[e.g.][]{andreon_low_2022}. Although robust, this method might suffer from incorrect galaxy member assignments due to projection effects. We extract from Magneticum the scaling relation between total stellar mass and $M_{200}$ and calibrate the halo mass proxy. 

    \item \textsc{Stellar luminosity of member galaxies.} Similarly to the stellar mass, the optical luminosity can be a good tracer of the stellar content of galaxy groups and clusters to derive the host halo mass. In observations, such scaling relation is particularly useful given that the optical luminosity is directly observable. We use the best-fit relation for the SDSS $r$-band listed in the Appendix Table of \cite{popesso_rass-sdss_2007} with a scatter $\sim$ 0.20 dex. 
\end{itemize}
Each scaling relation carries an intrinsic scatter $\sigma (\log M_{200}| Y)$ which is the scatter in recovering the logarithmic halo mass at a given proxy $Y$, generally as a function of the halo mass. Table~\ref{tab:scatter} illustrates the result for the sample of groups in LC30. 
\par
\begin{figure*}
    \centering
    \subfigure{\includegraphics[trim={0 1.4cm 0 0},clip, width=\textwidth]{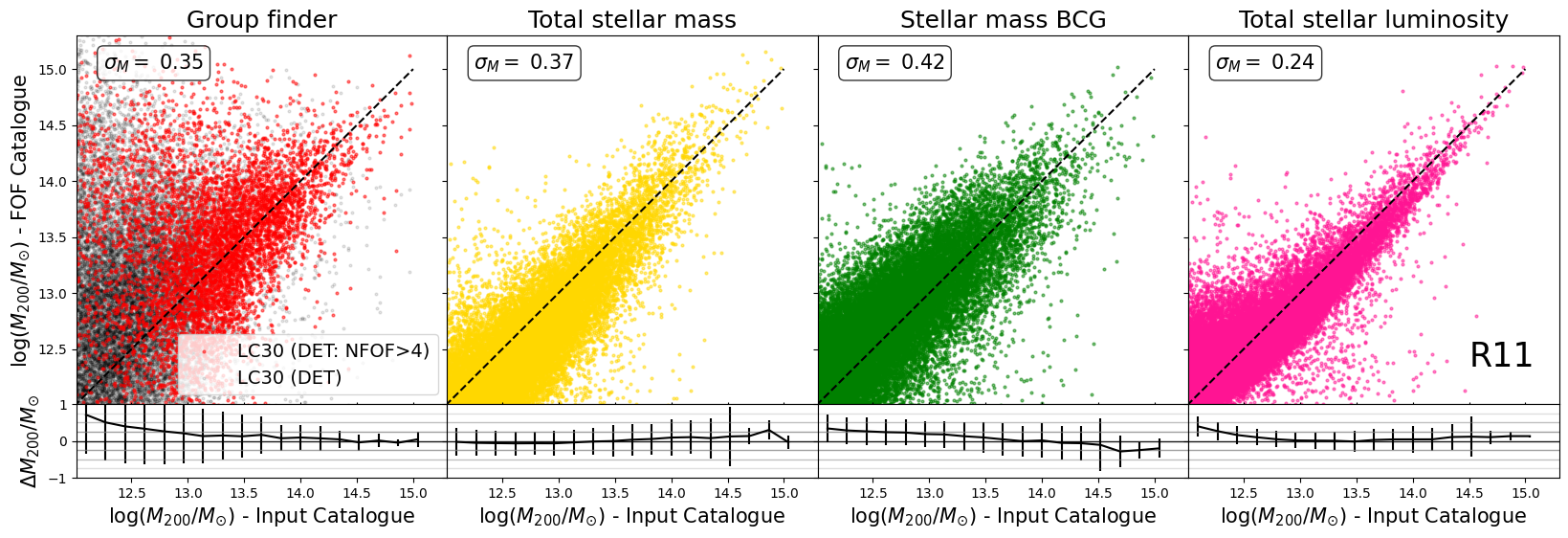}}     
    \subfigure{\includegraphics[trim={0 0 0 0.9cm},clip, width=\textwidth]{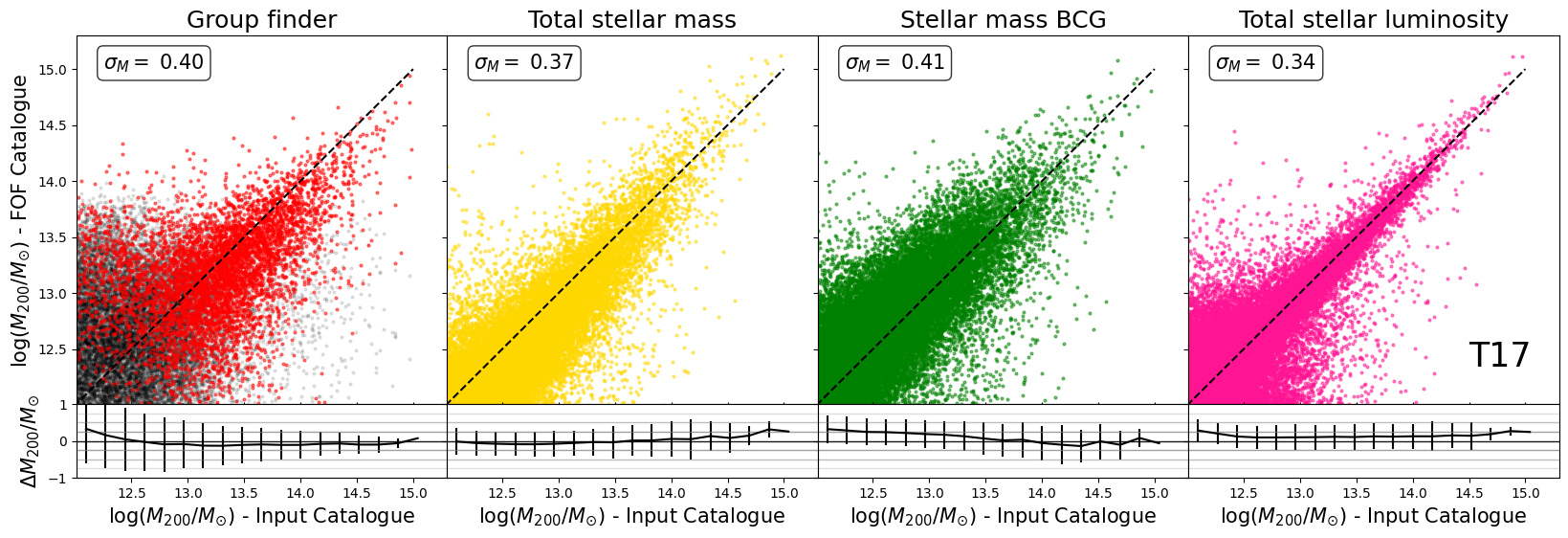}} 
    \subfigure{\includegraphics[trim={0 0 0 0.9cm},clip, width=\textwidth]{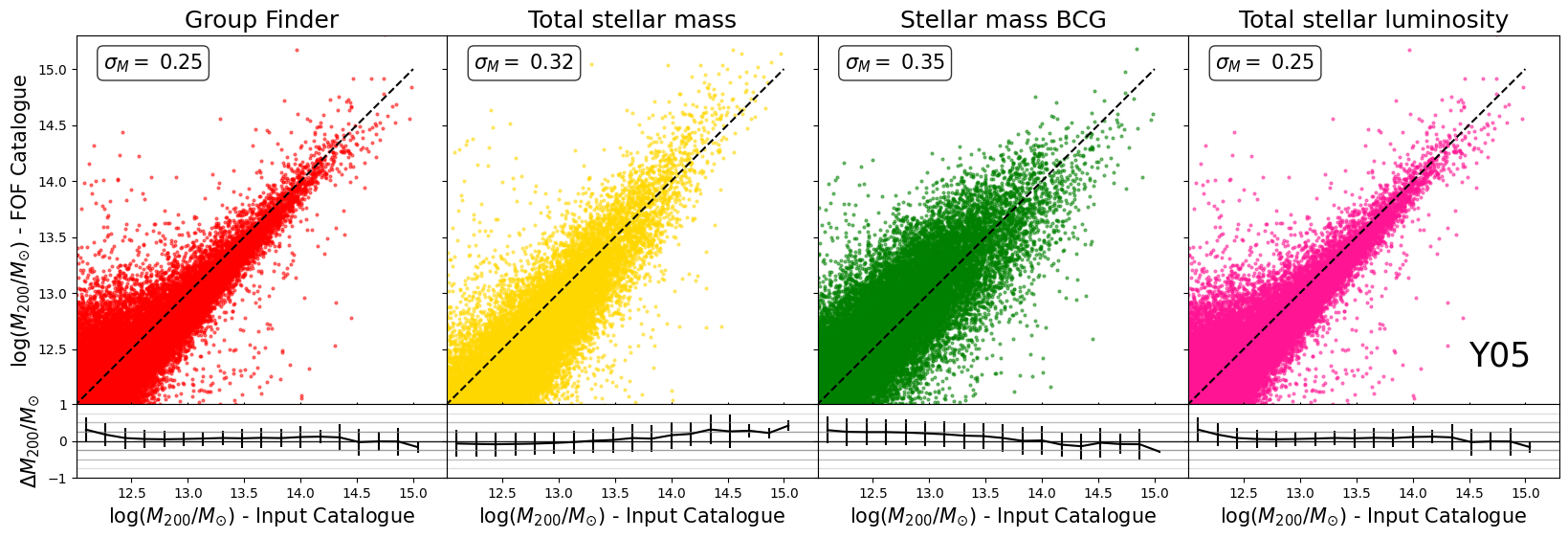}} 
    \caption{Comparison of the estimated halo mass in the FOF catalogues as a function of the true input mass from Magneticum. Each row represents the results from the different group finders: R11, T17, and Y05 from top to bottom. Each panel reports a different halo mass proxy. The algorithm developed by R11 and T17 estimates the halo mass based on the members' velocity dispersion, thus we show the distribution of points based on their richness (i.e. at least four members are the coloured points) in their panels. Points are the primary detections. Y05 uses the total stellar luminosity to probe the halo mass, thus there is no richness limit. Notice that in Y05 the first panel and last thus display the same data. For each distribution, we include a bottom rectangular panel with the scatter (median and standard deviation) in bins of the true Magneticum's halo mass. In the text at the top, we report the total median scatter (in dex). The dashed black line marks the 1:1 relation.  }
    \label{fig:MvM}
\end{figure*}

Fig.~\ref{fig:MvM} illustrates the scatter around the 1:1 relation (dashed black line) between the input halo mass and the matched halo mass estimated with different methods (reported at the top of each panel) and for each group finder (going from top to bottom R11, T17, and Y05). Considering that the group finders by R11 and T17 use the velocity dispersion of member galaxies to estimate the halo mass, we mark the data points with richness below five in grey in their leftmost panels. These estimates cannot be considered reliable with this mass proxy. Using the other methods gives us the advantage of keeping these halos since the other mass proxies are less affected. 
\par
Each panel reports in the lower section the relative scatter (median and standard deviation) of the primary detections as a function of the Magneticum's true halo masses, whereas the number enclosed within the parentheses in the text above reports the median scatter (in dex). Both metrics show that the stellar luminosity is the mass proxy with the smallest scatter, followed by the total stellar mass. 
\par
We can conclude two important results from the outcome of this experiment. Firstly, the stellar luminosity of the brightest member galaxies in a cluster or group scales fairly well with the host halo mass. This mass proxy is the best for our catalogue, closely followed by the total stellar mass. Secondly, using either one of these two proxies allows us to keep halos with multiplicity lower than five.

\subsection{Consistency of galaxy membership}
Another way to quantify the accuracy of the group finders is to investigate the consistency rate in recovering the underlying galaxy member population. In other words, after matching the halo catalogues, we cross-match their galaxy population and assess the impact of interlopers in the catalogue. Interlopers are the main contaminant in optical surveys -- affecting the purity of the galaxy members -- and they might cause significant damage when assessing the halo mass. Methods to recover the halo mass based on the purity of the galaxy member sample (e.g. velocity dispersion) will tend to under/overestimate the mass based on incorrect assignment.
\par
\begin{figure*}
    \centering
    \includegraphics[scale=0.55]{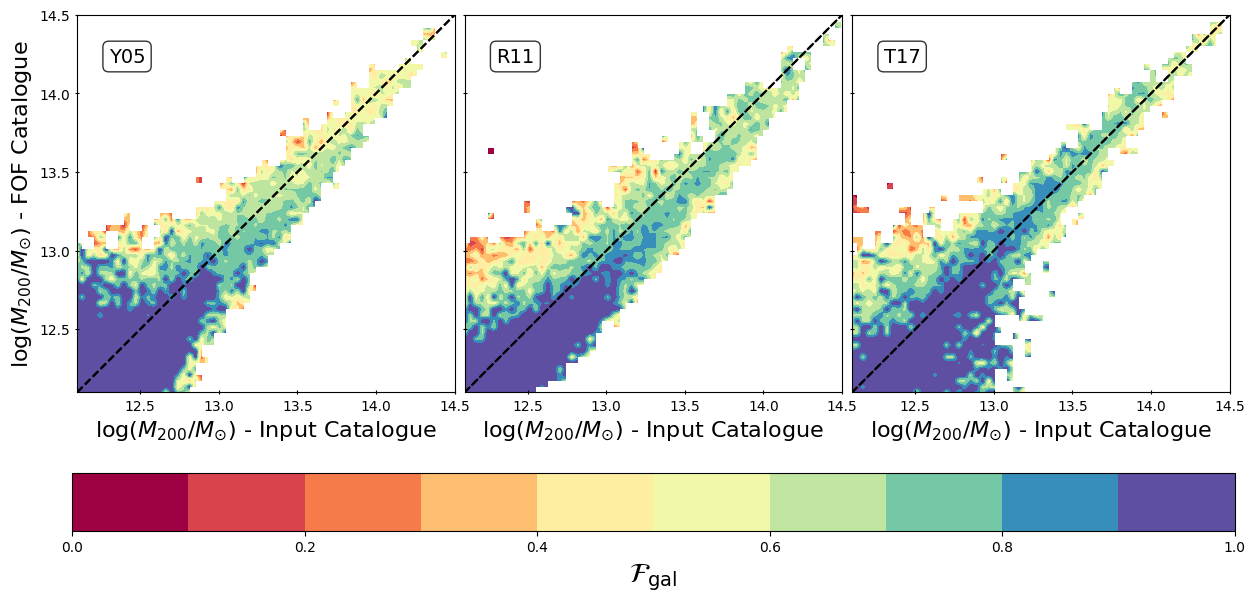}
    \caption{Scatter plot of the luminosity-based mass (best proxy) and true mass colour-coded for the accuracy of the membership $\mathcal{F}_{gal}$ for the three group finders. We define $\mathcal{F}_{gal}$ in Eq.~\ref{eq:F_gal}. }
    \label{fig:MvM_all_members}
\end{figure*}

Given our best mass proxy, how significant is the latent scatter in the recovered mass versus input mass as a function of galaxy membership? In other words, provided that FOF detection is matched to a true halo in Magneticum -- following the method in Sect.~\ref{sec:matching} -- how comparable are their two galaxy member populations?
We define $\mathcal{F_\mathrm{gal}}$ as
\begin{equation}
\label{eq:F_gal}
    \mathcal{F_\mathrm{gal}}= \frac{N_\mathrm{gal}^2 ( {\scriptstyle\mathrm{HALOS \,\cap\, FOF}} )}{N_\mathrm{gal} ({\scriptstyle\mathrm{HALOS}}) \,N_\mathrm{gal} ({\scriptstyle\mathrm{FOF}})},
\end{equation}
where $N_\mathrm{gal}( {\scriptstyle\mathrm{HALOS \,\cap\, FOF}} )$ is the number of true galaxy members, $N_\mathrm{gal} ({\scriptstyle\mathrm{HALOS}})$ and $N_\mathrm{gal} ({\scriptstyle\mathrm{FOF}})$ refer to the number of members in the Magneticum and FOF catalogues respectively. This quantity will be 1 if all the galaxy members are recovered in the matched catalogue and it will decrease otherwise. In Fig.~\ref{fig:MvM_all_members} we show the halo mass distribution estimated via the scaling relation with the stellar luminosity for all group finders and we colour-code the data according to $\mathcal{F}_\mathrm{gal}$. Overall, smaller $\mathcal{F}_\mathrm{gal}$ corresponds to an equivalently smaller deviation from the 1:1 relation. Low-mass halos ($< 10^{13} M_{\odot}$) generally have a high consistency, since only a few galaxies form them; however, the scatter in the mass in this range is quite large. At the groups and cluster scale, $\mathcal{F}_\mathrm{gal}$ decreases not necessarily hindering the capability of recovering the true halo mass. We find that correctly assigning the most massive satellites (i.e. contributing mostly to the stellar luminosity and mass budget) to the halo is far more important. 
\par
Given that the total stellar luminosity is the most accurate halo mass proxy, we might ask how comparable the member galaxy luminosity functions are: this is a good diagnostic of the distribution of galaxy properties within their DM halos between input and the FOF datasets. To address the issue, we study the conditional luminosity function \citep[CLF, hereafter;][]{yang_constraining_2003, van_den_bosch_towards_2003} which describes the average number of galaxies with luminosity within $L$ and $L+dL$ that reside in a halo of mass $M_{200}$. This formalism allows us to simultaneously address galaxies' clustering and abundance properties as a function of their luminosity.
\par
\begin{figure*}
    \centering
    \subfigure{\includegraphics[width=0.368\textwidth]{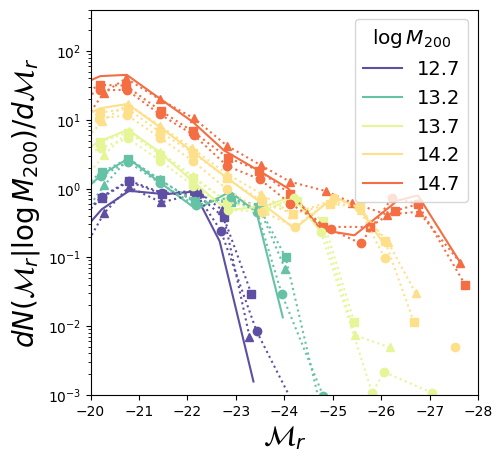}} 
    \subfigure{\includegraphics[width=0.31\textwidth, trim={2cm 0 0 0},clip]{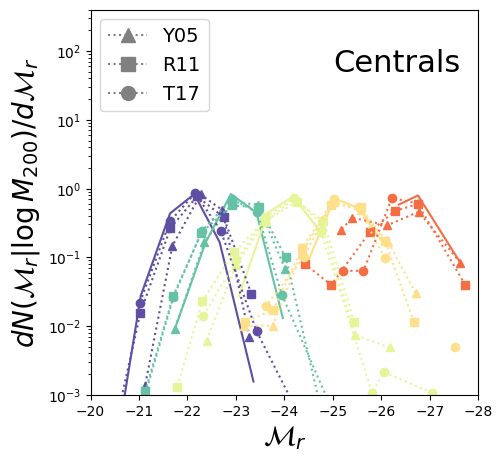}} 
    \subfigure{\includegraphics[width=0.31\textwidth, trim={2cm 0 0 0},clip]{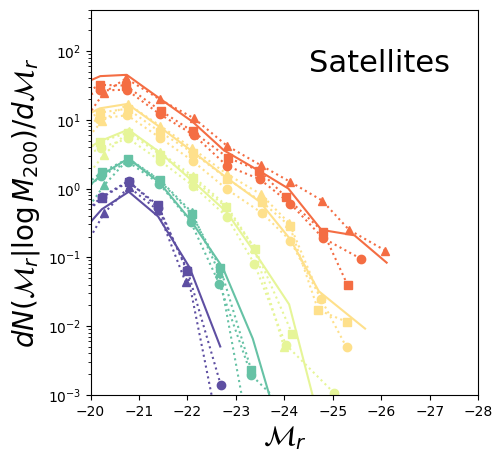}} 
    \caption{CLF of all the member galaxies (left panel), only central galaxies (central panel), and only satellites (right panel) as a function of halo mass. The halo masses are split into five equally spaced logarithmic bins, whose median is reported in the legend. The samples of each group finder are marked with symbols whose legend is listed in the central panel, the solid lines represent the input distribution from the SubFind halo catalogue.}  
    \label{fig:CLF}
\end{figure*}

We report the results in Fig.~\ref{fig:CLF}, expressing the CLF in terms of the rest-frame magnitude in the SDSS $r$-band $\mathcal{M}_r$. The bright end of the CLF extends to very low values of $\mathcal{M}_r$, which is consistent with a broad GSMF (as discussed in Fig.~\ref{fig:GSMF}) and no dust attenuation in the stellar luminosity. Such CLF is not comparable with observational data, however, it provides us with insights into the recovery rate of the group finders, since we only ask them to be self-consistent with the simulations. Furthermore, it is useful to disentangle the role played by the central galaxies (central panel) and the satellites (right panel) in the total CLF (left panel) since by definition the central galaxy is the brightest galaxy in each DM halo. We split the groups into five halo mass bins and report the median halo mass in the legend. The solid lines represent the results from the Magneticum halo catalogue, whereas the dotted lines mark the different FOF catalogues, according to the various symbols. We check that including FRAG\_2 has a negligible effect on the plots, and only at the low-luminous end, and therefore we do not include them in the count.
\par
We do not observe any significant difference between the input and FOF distributions, except in the highest mass bin where we see that Y05 tends to overestimate the intermediate luminous satellite population (i.e. in the magnitude range between $-24$ and $-26$) as opposed to the other CLFs. This result agrees with the decrease in the completeness at high masses seen in Fig.~\ref{fig:completeness_contamination} when accounting for the galaxy membership. Furthermore, splitting the sample between the central and satellite galaxies allows us to disentangle the two peaks in the total CLF in most mass bins. Overall, the group finders provide reliable distribution of the galaxy properties, self-consistent within the simulation. This result demonstrates that the group finders can also offer reliable catalogues of galaxies within different environments. 

\section{Discussion: Halo mass distribution}
\label{sec:hidden_bias}
Having discussed the completeness and contamination of each optical catalogue and derived the best halo mass proxy, we can now investigate how representative the recovered groups and clusters catalogues are of the underlying (true) halo population in LC30.
\par
To further expand the discussion, we include the recovered halo mass distribution extracted in \cite{marini_detecting_2024} from the same simulated lightcone. \cite{marini_detecting_2024} created a full mock X-ray observation in LC30 as the extended ROentgen Survey with an Imaging Telescope Array (eROSITA) would deliver. The real survey will scan the entire X-ray sky both in the soft band (i.e. $0.2-2.0$ keV) and hard (i.e. $2.0-10$ keV) allowing to probe galaxy clusters and groups at these wavelengths in an unprecedented volume. The mock observation is performed with an exposure time similar to the 4 years eROSITA All-Sky Survey (eRASS:4). The authors model self-consistently the X-ray photons of the hot gas, AGNs, and XRB using \texttt{PHOX} \citep{biffi_observing_2012, biffi_investigating_2013, biffi_agn_2018, vladutescu-zopp_decomposition_2023} which are post-processed by SIXTE \citep{dauser_sixte_2019}, the official end-to-end eROSITA simulator, to account for the telescope's technical features and scanning strategy. Following the spectral decomposition of the eFEDS background emission in \cite{liu_erosita_2022}, foreground and background components are added. Extended and point source detections are derived by running the eROSITA Science Analysis Software System (eSASS) on the event files. Each extended detection is cross-matched to the input halo catalogue in a similar manner to what is described in Sect.~\ref{sec:matching}. From the X-ray catalogue, the authors derive the expected completeness and contamination of the sample, and the luminosity function and discuss the intrinsic systematics included in such observation. 
\par
\begin{figure}
    \centering
    \includegraphics[width=0.95\linewidth]{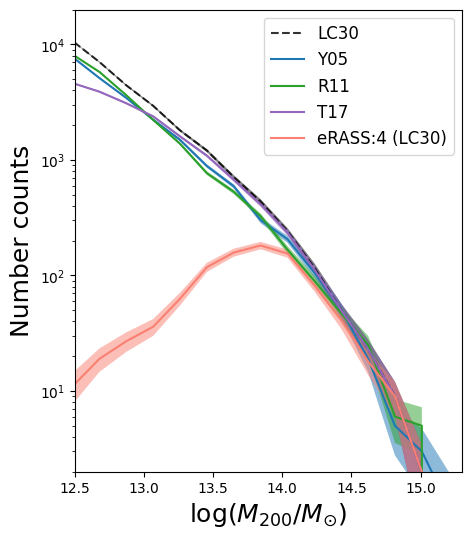}
    \caption{Number count distribution of the halo masses in the input catalogue (black dashed line) and the other samples. We report the Poissonian uncertainty with the shaded area. }
    \label{fig:N_counts}
\end{figure}

The comparison between X-ray and optical catalogues allows one to examine different selection effects at the groups and clusters scale. With this approach in mind, in Fig.~\ref{fig:N_counts} we present the complete halo mass distribution in LC30 (dashed black line), from the optical surveys (i.e. Y05, R11, and T17) and the X-ray catalogue. \cite{marini_detecting_2024} thoroughly discuss the selection effects in the X-ray sample related to the amount of X-ray emitting hot gas in the halos. Such bias leads to the detection of gas-rich and low-entropy systems at fixed halo mass, preferentially. These are the vast majority for $M_{200}>10^{14} M_{\odot}$ and diminish for smaller masses. This effect is evident in our catalogues and causes a drastic decrease of detected halos for $M_{200}<10^{14} M_{\odot}$. On the other hand, the optical catalogues (using the halo mass derived from the stellar luminosity, and thus allowing us to include objects with low multiplicity) tend to be more complete for lower masses. Overall, the T17 catalogue provides the most accurate halo mass distribution.
\par
We argue that the larger statistics offered by the optical selection allow one to study halos at the galaxy group scales with far fewer biases than in X-ray selection. In this context, stacking X-ray observations on optically selected halos will allow mapping the distribution of baryonic mass at smaller scales and potentially infer the average properties of the hot gas in galaxy groups. A number of studies have exploited this in the group's regime, and without the aim to be complete, we name a few: \cite{crossett_xxl_2022, ota_erosita_2023, popesso_x-ray_2024}. 

\section{Summary}
\label{sec:conclusions}
The results presented in this study are based on predictions from the Magneticum Pathfinder, a suite of cosmological hydrodynamical simulations. The parent \textit{Box2/hr} simulation serves to extract a mock spectroscopic catalogue of optical galaxy groups and clusters down to \(z<0.2\), complete for stellar masses \(\geq10^{9.8} M_{\odot}\). Galaxies in the lightcone LC30 (with a field of view of \(30 \times 30\) deg\(^2\)) are simulated with observed magnitudes in the SDSS bandpass (i.e. \emph{u, v, g, r}, and \emph{i}) and an additional $K$-correction. The corresponding group catalogues are constructed using three widely employed group finders \citep[i.e.][]{robotham_galaxy_2011, yang_halo-based_2005, tempel_merging_2017} on the mock galaxy catalogues. We compare the results in terms of completeness, contamination, and recovery of the halo mass.

Our main findings are summarised as follows:
\begin{itemize}
    \item The optical catalogues show that at least 80 per cent of the halos with \(M_{200}>10^{13} M_{\odot}\) are recovered (Fig.~\ref{fig:completeness_contamination}). Contamination at the low-mass end (that is, $M_{200}<10^{13} M_{\odot}$) is due to projection effects and fragmentation, since many FOF groups correspond to FRAG\_2 of massive halos in all group finders. Primary fragments (i.e. FRAG\_1) have no redshift dependence, albeit a significant halo mass dependence.
    \item Above $M_{200}>10^{13} M_{\odot}$, the FOF catalogue is also accurate ($\sim 80$\%) in assigning the galaxy members to their respective group, as shown in Fig.~\ref{eq:completness}. The true and recovered CLFs for centrals and satellites are broadly consistent at all masses, allowing us to further confirm that these group finders are also robust for studies of galaxies as a function of the environment. 
    \item Fragmentation and incorrect galaxy membership might systematically bias the halo mass estimates by the group finders. This is crucial for group finders that use the velocity dispersion of the galaxy members as a mass proxy via dynamical modelling. We used total stellar mass, central galaxy stellar mass, and $r$-band luminosity to determine the best halo mass proxy. Stellar luminosity shows the smallest scatter for all three group finders (\ 0.24–0.40\ dex), followed by the total stellar mass (\ 0.37–0.49\ dex).
    \item The recovery rate of the BCG in all groups is also very high (refer to Fig.~\ref{fig:BCG_correct}), allowing the calibration of halo masses on the BCG stellar mass with a small scatter.
    \item Using stellar luminosity (or total stellar mass) as a halo mass proxy allows us to estimate halo masses with sufficient accuracy, even when not all member galaxies are correctly associated. This holds if at least some of the most luminous (or most massive) member galaxies are included, a common scenario for all group finders.
    \item The halo mass distribution in Fig.~\ref{fig:N_counts} highlights the selection effect in X-ray surveys of galaxy groups. While optical group finders may require tuning for halo mass calibration, they are powerful tools for discovering and detecting galaxy groups, even in the local Universe.
\end{itemize}
Our study underscores the suitability of optical (spectroscopic) surveys for detecting galaxy clusters and groups in the local Universe. Interlopers and fragmentation, primarily affecting high-mass halos, can impact our ability to use galaxy members to probe halo masses. However, different mass proxies can mitigate this negative impact and yield complete halo catalogues. Ultimately, the error budget for determining accurate halo masses should account for all possible systematics that cannot be further improved through simulation tuning. A clear understanding of both selection effects in optical data and irreducible systematics in group finders is essential for highly complete, optically selected samples of galaxy groups.

\begin{acknowledgements}
       We thank the referee for the valuable suggestions to the paper. 
      This project has received funding from the European Research Council (ERC) under the European Union’s Horizon Europe research and innovation programme ERC CoG (Grant agreement No. 101045437). KD acknowledges support by the COMPLEX project from the European Research Council (ERC) under the European Union’s Horizon 2020 research and innovation program grant agreement ERC-2019-AdG 882679. The calculations for the Magneticum simulations were carried out at the Leibniz Supercomputer Center (LRZ) under the project pr83li. ET acknowledges the ETAg grant PRG1006 and the CoE project TK202, which was funded by the HTM. XY acknowledges the support from the National Key R\&D Program of China (2023YFA1607800, 2023YFA1607804). NM acknowledges funding by the European Union through a Marie Sk{\l}odowska-Curie Action Postdoctoral Fellowship (Grant Agreement: 101061448, project: MEMORY). However, the views and opinions expressed are those of the author alone and do not necessarily reflect those of the European Union or the Research Executive Agency. Neither the European Union nor the granting authority can be held responsible for them. MB acknowledges the support of McMaster University through the William and Caroline Herschel Fellowship.
      This project has received funding from the European Research Council (ERC) under the European Union’s Horizon Europe research and innovation programme ERC CoG (Grant agreement No. 101045437). KD acknowledges support by the COMPLEX project from the European Research Council (ERC) under the European Union’s Horizon 2020 research and innovation program grant agreement ERC-2019-AdG 882679. The calculations for the Magneticum simulations were carried out at the Leibniz Supercomputer Center (LRZ) under the project pr83li. ET acknowledges the ETAg grant PRG1006 and the CoE project TK202, which was funded by the HTM. XY acknowledges the support from the National Key R\&D Program of China (2023YFA1607800, 2023YFA1607804). NM acknowledges funding by the European Union through a Marie Sk{\l}odowska-Curie Action Postdoctoral Fellowship (Grant Agreement: 101061448, project: MEMORY). However, the views and opinions expressed are those of the author alone and do not necessarily reflect those of the European Union or the Research Executive Agency. Neither the European Union nor the granting authority can be held responsible for them. MB acknowledges the support of McMaster University through the William and Caroline Herschel Fellowship.
      This project has received funding from the European Research Council (ERC) under the European Union’s Horizon Europe research and innovation programme ERC CoG (Grant agreement No. 101045437). KD acknowledges support by the COMPLEX project from the European Research Council (ERC) under the European Union’s Horizon 2020 research and innovation program grant agreement ERC-2019-AdG 882679. The calculations for the Magneticum simulations were carried out at the Leibniz Supercomputer Center (LRZ) under the project pr83li. ET was supported by the Estonian Ministry of Education and Research (grant TK202), Estonian Research Council grant (PRG1006) and the European Union's Horizon Europe research and innovation programme (EXCOSM, grant No. 101159513). XY acknowledges the support from the National Key R\&D Program of China (2023YFA1607800, 2023YFA1607804). NM acknowledges funding by the European Union through a Marie Sk{\l}odowska-Curie Action Postdoctoral Fellowship (Grant Agreement: 101061448, project: MEMORY). However, the views and opinions expressed are those of the author alone and do not necessarily reflect those of the European Union or the Research Executive Agency. Neither the European Union nor the granting authority can be held responsible for them. MB acknowledges the support of McMaster University through the William and Caroline Herschel Fellowship.
\end{acknowledgements}

%
    \bibliographystyle{aa} 
   \bibliography{references} 
%
\begin{appendix}
\section{Testing the GSMF with UniverseMachine}
\label{sec:UniverseMachine}
Some concerns may be raised regarding the limitations of these tests being based on a single simulation. For example, the shape of the GSMF in Fig.~\ref{fig:GSMF} -- particularly the presence of very massive central galaxies within their host halo mass -- might influence the mass proxy derived from the central galaxy.
\par
Here, we benchmark our predictions by re-running the detection and matching procedure outlined in Sect.~\ref{sec:group_finders} in a lightcone generated with UniverseMachine \citep{behroozi_universemachine_2019}. The geometrical configuration is the same (i.e. $30\times 30$ deg$^{2}$ down to $z<0.2$ and complete down to $M_{\star}\geq10^{9.8} M_{\odot}$), however, the parent simulation Bolshoi-Planck \citep{klypin_multidark_2016} is a DM-only cosmological box 250 $h^{-1}$ Mpc on a side with 2048$^3$ particles which uniquely integrates empirical data to create a detailed and comprehensive representation of galaxy population. The subhalos are identified using ROCKSTAR \citep{behroozi_rockstar_2013}
\par
UniverseMachine models individual galaxies' star formation rates (SFRs) within their DM halos, by parametrising the maximum circular velocity at peak historic halo mass, the assembly history, and redshift. Parameters describing the SFR and the fraction of quenched galaxies based on these variables are constrained using a variety of observations spanning a wide redshift range (up to $z < 10$). These observations include galaxy abundances, the cosmic SFR, the fraction of quenched galaxies as a function of galaxy observables, and the auto- and cross-correlation of star-forming and quenched galaxies. As a result, the UniverseMachine produces numerous predictions, including the stellar-to-halo mass relation as a function of redshift, the star formation histories of galaxies, their correlation functions, and statistics on infall and quenching for satellites \citep{behroozi_universemachine_2023}. \par
A mock galaxy catalogue generated with this method is constrained by galaxies’ observed stellar mass functions, SFRs (specific and cosmic), quenched fractions, UV luminosity functions, UV–stellar mass relations, IRXUVrelations, auto- and cross-correlation functions (including quenched and star-forming subsamples), and quenching dependence on environment. Although this simulation lacks modelling of the optical emission, the group finders can rely on the 3D spatial properties and masses of galaxies to reconstruct the halo catalogue. Therefore, we test our second-best halo mass proxy (i.e. the stellar mass of the member galaxies) against the group finder results and discuss the catalogue's completeness and contamination. 
\par
\begin{figure*}
    \centering   \subfigure{\includegraphics[width=0.44\textwidth]{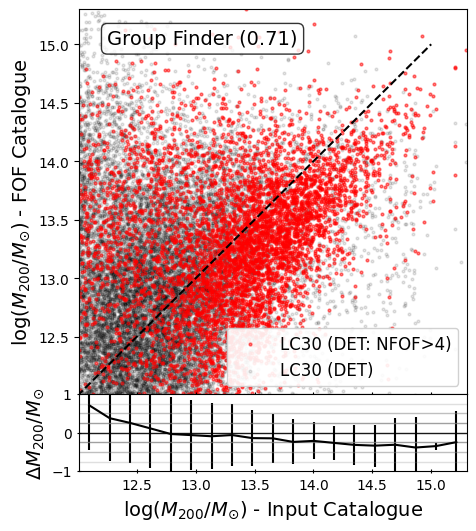}}     \subfigure{\includegraphics[width=0.44\textwidth]{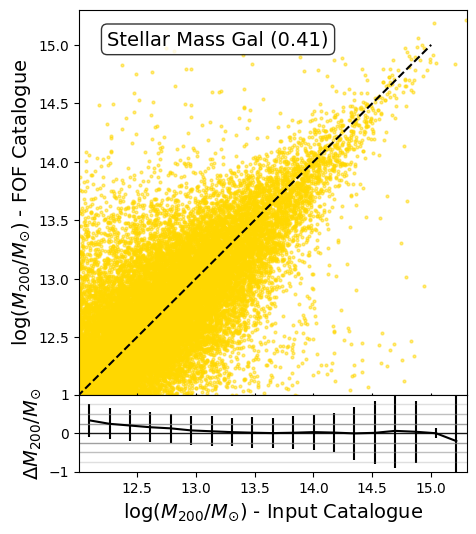}}
    \caption{Scatter plot of the halo mass distribution when using the group finder's method in R11 (i.e. the member galaxies' velocity dispersion; left panel ) and the stellar mass (right panel). The legend is the same as in Fig.~\ref{fig:MvM}.  }
    \label{fig:MvM_R_UM}
\end{figure*}

Fig.~\ref{fig:MvM_R_UM} illustrates the recovery of the halo mass when using the velocity dispersion (left panel) or the stellar mass of the member galaxies (right panel) for the halo catalogue derived with R11. We use the same legend as in Fig.~\ref{fig:MvM} to provide an easy comparison between the two. The scatter is larger (i.e. 0.71) than when using the group finder with Magneticum (0.35), nevertheless when using the stellar mass as a mass proxy, it becomes comparable. We argue that the systematically higher stellar mass function observed in Magneticum does not imply a better recovery of the halo mass when using the stellar mass as a proxy. 
\par
\begin{figure}
    \centering
    \includegraphics[scale=0.75]{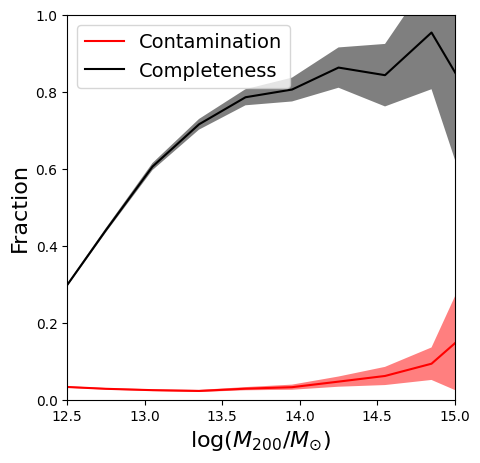}
    \caption{Completeness and contamination for the input halo mass from the UniverseMachine mock catalogue run with R11. The shaded bands mark the dispersion given by the Poissonian error. }
    \label{fig:completeness_contamination_UM}
\end{figure}

Fig.~\ref{fig:completeness_contamination_UM} displays the completeness and contamination levels found for the halo catalogue. Completeness and contamination confirm our previous results for galaxy groups and clusters.

\end{appendix}
\end{document}